\documentclass[12pt]{article}
\usepackage[pctex32]{graphics}
\begin{document}
\vspace{2cm}
\begin{center}
{\bf  \Large  Quantum Field Effects on Cosmological Phase Transition in
Anisotropic Spacetimes}
\vspace{1cm}

                      Wung-Hong Huang\\
                       Department of Physics\\
                       National Cheng Kung University\\
                       Tainan,70101,Taiwan\\

\end{center}
\vspace{2cm}
   The one-loop renormalized effective potentials for the massive $\phi^4$ theory on the spatially homogeneous models of Bianchi type I and Kantowski-Sachs type are evaluated.   It is used to see how the quantum field affects the cosmological phase transition in the anisotropic spacetimes.  For reasons of the mathematical technique it is assumed that the spacetimes are slowly varying or have specially metric forms. We obtain the analytic results and present detailed discussions about the quantum field corrections to the symmetry breaking or symmetry restoration in the model spacetimes.
\vspace{3cm}
\begin{flushleft}
E-mail:  whhwung@mail.ncku.edu.tw\\
Classical and Quantum Gravity Vol.10 (1993) 2021-2033
\end{flushleft}

\newpage
\section{Introduction}
    Quantum fields have a profound influence on the dynamical behavior of the early  universe [1].  Among them, the quantum-field effects on the cosmological phase transition have been investigated by many authors [2-14]. From analyses based on the renormalized effective potential it was concluded that the scalar-gravitational coupling $\xi$  and the magnitude of the scalar curvature R crucially determine the fate of symmetry.

   The effect of anisotropy on the static spacetimes like Mixmaster or Taub universe on the process of symmetry breaking and restoration has been discussed [4, 5]. We have recently investigated the symmetry breaking in a Bianchi type I spacetime (which is assumed to be slowly varying), the inhomogeneous spacetime (which is assumed to be slowly varying) and a rotational spacetime (with the Godel metric) respectively [12-14], and this paper becomes the fourth of the series.

   The purpose of this paper is to present our further investigations about the quantum-field effect on the cosmological phase transition in anisotropic spacetimes. We will evaluate the one-loop renormalized effective potentials for the massive $\phi^4$ theory on  the spatially homogeneous models of Bianchi type I and Kantowski-Sachs type. To enable the analytic results to be found we assume that the spacetime metric is either slowly varying or has a specified form. Using the results we discuss the properties of the quantum-field corrections to the symmetry breaking or symmetry restoration in these two model spacetimes respectively.

   There are two reasons that lead us to re-investigate the problem of the quantum field effect on the cosmological phase in anisotropic spacetimes. First, it is not clear [9] that the systems evolving in a spacetime with large shear Q do not invalidate the slowly varying background field assumption in the treatment of symmetry behavior in dynamical spacetimes. We thus hope to have an anisotropic spacetime in which the one-loop renormalized effective potential can be evaluated exactly without using the adiabatic approximation, and thus the quantum field effect on the fate of symmetry could be known in the affirmative. This motivates us to study the model on a specified metric form. (Note that it is mathematically difficult to find the exact mode function to evaluate the effective potential, even in a generally isotropic spacetime.) Second, we want to know how general the properties found in the Bianchi type I are in other anisotropic spacetimes. This motivates us to study the model on the Kantowski-Sachs
spacetime.

   This paper is organized as follows. In section 2, the method and basic formulae are introduced. In section 3 we evaluated the one-loop renormalized effective potential for the $\phi^4$ theory in the Bianchi type I spacetime. The cases of a model spacetime which either is slowly varying or has a specified metric are evaluated respectively.   Using the results we discuss the properties of the quantum field corrections to the symmetry breaking or symmetry restoration. The evaluation and analyses on the Kantowski-Sachs spacetime are given in section 4. Section 5 is devoted to discussion.

\section {Formalism}
\subsection{Lagrangian and effective potential}

The Lagrangian density describing a massive (m), self-interacting ($\lambda$) scalar field ($\phi$) coupled arbitrarily ($\xi$) to the gravitational background is

$$ L= \sqrt {-g} \left({1\over2}\left[g^{\mu\nu}\partial_\mu \phi \partial_\nu \phi - (m^2 + \xi R)\phi^2\right] -{\lambda\over4!} \phi^4\right).  \eqno{(2.1)}$$
\\
The equation of motion associated with it is

$$ g^{\mu\nu}\nabla_\mu\nabla_\nu \phi +(m^2+\xi R)\phi  +{\lambda\over 3!}\phi ^3 = 0.  \eqno{(2.2)}$$
Let us write
$$ \phi =\phi _c + \phi _q, ~~~~~~~~~~~~<\phi _c> = 0,\eqno{(2.3)}$$
where $\phi _c$  is a classical field and $\phi_q$ is a quantum field with a vanishing vacuum expectation value; then, taking the expectation value of (2.2) and introducing the renormalized parameters $m^2_r$, $\xi_r$ and $\lambda_r$, by

$$m^2 = m^2_r + \delta m^2 , ~~ \xi= \xi_r + \delta \xi, ~~~ \lambda_r = \lambda + \delta \lambda, \eqno{(2.4)}$$
\\
we have the field equation for the classical field $\phi_c$ 

$$  g^{\mu\nu}\nabla_\mu\nabla_\nu \phi_c + [(m^2+\delta m^2)+(\xi +\delta \xi)R]\phi_c + {\lambda + \delta \lambda\over 2}<\phi^2_q> \phi_c +  {\lambda + \delta \lambda\over 3!} \phi_c^3 =0 .\eqno{(2.5)}$$
\\
and, to one-loop quantum effect, the field equation for the quantum field $\phi_q$ [3]

$$ g^{\mu\nu}\nabla_\mu\nabla_\nu \phi_q +[m_r^2+\xi_r R]\phi_q  +{\lambda_r\over 2}\phi^2_c \phi_q=0. \eqno{(2.6)}$$
\\
The effective potential $V_{eff}$ is given by

$$ {\partial V_{eff}\over \partial \phi_c} = [(m^2+\delta m^2)+(\xi +\delta \xi)R]\phi_c + {\lambda + \delta \lambda\over 2}<\phi^2_q> \phi_c +  {\lambda + \delta \lambda\over 3!} \phi_c^3\eqno{(2.7)}$$
\\
where the value of $<\phi_q^2>$  is formally divergent and we can use the following renormalization conditions to make it finite [1-3]

$$m^2_r = {\partial^2 V_{eff}\over \partial \phi_c^2}|_{\phi_c=R=0}, ~~~\xi_r = {\partial^3 V_{eff}\over \partial R \partial \phi_c^2}|_{\phi_c=R=0}, ~~~\lambda_r = {\partial^4 V_{eff}\over \partial^4 \phi_c^2}|_{\phi_c=R=0}, ~~~
\eqno{(2.8)}$$
\\
The above conditions will give the counter terms, $\delta m^2$, $\delta\xi$  and $\delta \lambda$ defined in equation (2.4).

\subsection {Quantization}

To evaluate $<\phi_q^2>$ we adopt the canonical quantization relations

$$ [\phi_q(t,x),\phi_q(t,y)] = [\pi_q(t,x),\pi_q(t,y)]=0, ~~~ [\phi_q(t,x), \pi_q(t,y)]= i \delta^2 (x-y), \eqno{(2.9)}$$
\\
where the conjugate momentum $\pi_q(t,x)$ is defined by

$$ \pi_q = \partial~ L/\partial (\partial_t\phi).  \eqno{(2.10)}$$
\\
Due to the space homogeneity we can expand the quantum field $\phi_q $ by the summation or integration over modes in the form

$$\phi_q(t,x) = C^{-1/2}(t)\int d\mu(k)[a_k ~\chi_k (t) Y_k(x) + a^\dagger_k~ \chi^*_k (t) Y^*_k(x)], \eqno{(2.11)}$$
\\
where $Y_k(x)$ is a normalized eigenfunction of the spatial-part field equation, while $\chi_ k(t)$  is that of the time part. We symbolically denote the eigenvalues of the special-part wave equation by $k$ which may be continuous or discrete. We also symbolized the summation or integration thereby involved by a measure $\mu(k)$.  Note that the factor $C^{-1/2}$ is usually introduced to simplify the time part of the field equation. The field equation described the mode function $ \chi_k (t)$ can be found by substituting equation (2.11) into field equation (2.6).

   The creation and annihilation operations $a_k$ and $a^\dagger_k$ will satisfy the usual relations.
$$ [a_k,a_l]=[a^\dagger_k,a^\dagger_l]=0,~~~ [a^\dagger_k, a^\dagger_l] = \delta^3(k-l)\eqno{(2.12)}$$
provided that the mode function $\chi_k$ satisfies the Wronskian condition
$$ \chi_k \dot\chi^*_k- \chi^*_k \dot\chi_k = i. \eqno{(2.13)}$$
From the above method we then obtain
$$ <\phi^2_q(t,x)> = C^{-1}(t) \int d\mu(k) \chi_k (t)\chi^*_k (t) Y_k(x) Y^*_k(x). \eqno{(2.14)}$$
Note that, as the spacetime we consider is homogeneous the value of $ <\phi^2_q(t,x)>$ will eventually be spatial-independent although the above equation appears to be x dependent.

    To proceed we must have an explicit function form of the mode solutions $\chi_k(t)$ and $Y_k(x)$, which can only be found after specifying the background spacetime.

\section {$\phi^4$ on Bianchi type I spacetime}

We first consider the 1 + 3 dimensional Bianchi type I spacetime which has the line element

$$ ds^2 = C(\eta) d \eta ^2 - a_1^2(\eta) dx^2 - a_2^2(\eta) dy^2 - a_3^2(\eta) dz^2 , ~~~~ C = (a_1 a_2 a_3)^{2/3}.\eqno{(3.1)}$$
\\
In this model spacetime the mode function and measure defined in equation (2.11) become

$$ Y_k(x) = (2\pi)^{-1/2} exp(i k x), \eqno{(3.2)}$$
$$ \int d\mu(k) = \int d^3 k,  \eqno{(3.3)}$$
\\
and equation (2.14) becomes
$$ <\phi^2_q(\eta)> = {1\over 8\pi C(\eta)} \int d^3 k \chi_k (\eta) \chi^*_k (\eta).\eqno{(3.4)}$$
In the above equation the time-part mode function $\chi_k$ satisfies the wave equation

$$ \ddot \chi_k +\left[C(\eta) \left[m^2_r + (\xi_r-{1\over 6})R +{\lambda_r\over2}\phi_c^2+\sum_i {k_i^2\over  a^2_i}\right] + Q(\eta)\right] \chi_k =0, \eqno{(3.5)}$$
\\
where the spacetime curvature function R and anisotropic function Q are

$$ R = 6C^{-1}(\dot H +H^2 +Q) ,~~~~ H= \sum_i h_i,~~~~~ h_i = \dot a_i/a_i, \eqno{(3.6)}$$
$$ Q = {1\over 32}\sum_{i<j}(h_i-h_j)^2. \eqno{(3.7)}$$
\\
   To obtain the explicit form of $\chi_k$ we consider the following two cases.

\subsection { WKB Approximation}

When the metric is slowly varying equation (3.5) possesses the WKB approximation solution

$$ \chi_k = (2W_k)^{-1/2} exp\left( -i\int d \eta ~W_k\right),\eqno{(3.8)}$$
\\
where $W_k$ can be approximated by [15-17]
$$W_k = \left( C(\eta)\left[m^2_r + (\xi_r-{1\over 6})R +{\lambda_r\over2}\phi_c^2+\sum_i {k_i^2\over  a^2_i}\right]  + Q \right) ^{1/2}\eqno{(3.8)}$$

Substituting the above solutions into equation (3.4) we have

$$<\phi_q^2> = {1\over 16\pi^3 C} \int d^3k   \left( C(\eta)\left[m^2_r + (\xi_r-{1\over 6})R +{\lambda_r\over2}\phi_c^2+\sum_i {k_i^2\over  a^2_i}\right] +Q\right) ^{-1/2}$$
$$={1\over 16\pi^3 } \int d^3k   \left[m^2_r + (\xi_r-{1\over 6})R +{\lambda_r\over2}\phi_c^2+\sum_i {k_i^2}+ {Q\over C}\right]  ^{-1/2}$$
$$={1\over 4\pi^2 } \int_0^\Lambda dk k^2  \left[m^2_r + (\xi_r-{1\over 6})R +{\lambda_r\over2}\phi_c^2+ k^2+ {Q\over C}\right]  ^{-1/2}$$
$$={1\over 8\pi^2 } [\Lambda^2 + {1\over2} \left(m^2_r + (\xi_r-{1\over 6})R +{\lambda_r\over2}\phi_c^2+{Q\over C}\right) \hspace{1.5cm}$$
$$\times \left(1+ ln [(m^2_r + (\xi_r-{1\over 6})R +{\lambda_r\over2}\phi_c^2+{Q\over C})/ 4\Lambda^2] \right)],\eqno{(3.10)}$$
\\
where we have introduced a momentum cut-off $\Lambda$  to regularize the k integration. From the renormalization conditions of equation (2.8) we can evaluate the renormalization counterterms

$$\delta m^2 = -{\lambda_r\over 16\pi^2}\left [ \Lambda^2 + {1\over2} \left(m^2_r + {Q\over C}\right)\left(1+ ln [(m^2_r +{Q\over C})/ 4\Lambda^2] \right) \right ] \eqno{(3.11)}$$
$$ \delta \xi = -{\lambda_r\over 16\pi^2}(\xi_r -{1\over6})\left(1+ {1\over2}ln [(m^2_r +{Q\over C})/ 4\Lambda^2] \right) ,\eqno{(3.12)}$$
$$ \delta \lambda  = -{\lambda_r^2 \over 16\pi^2}\left(3+ {3\over2}ln [(m^2_r +{Q\over C})/ 4\Lambda^2] \right) ,\eqno{(3.13)}$$
\\
Substituting these results into equation (2.7) we finally obtain the derivative of the one-loop renormalized effective potential

$$ {\partial V_{eff} \over \partial\phi_c} = m^2_r\phi_c + \left[\xi_r - {\lambda_r\over32\pi^2} (\xi_r-{1\over 6})\right] R \phi_c + \left( {\lambda_r\over3!}-{\lambda_r^2\over 64\pi^2}\right)\phi_c^3 \hspace{5cm}$$
$$+{\lambda_r\phi_c\over 32\pi^2}  \left[\left(m^2_r + (\xi_r-{1\over 6})R +{\lambda_r\over2}\phi_c^2+{Q\over C}\right) \right] ln \left[m^2_r + (\xi_r-{1\over 6})R +{\lambda_r\over2}\phi_c^2+{Q\over C}\over m^2_r + Q/C \right] \eqno{(3.14)}$$
\\
Note that once we let the anisotropy in the above equation be zero our result is
consistent with that in the symmetric homogeneous case [10].

   We are now 1n a position to discuss the quantum field effects on the cosmological phase transition. Taking the derivation of the above equation, then in the limit $\phi_c\rightarrow 0$ we obtain the effective mass

$$ m^2_{eff} = (m_r^2 +\xi_r R) + {\lambda_r\phi_c\over 32\pi^2} [-(\xi_r-{1\over6})R + \left(m^2_r + (\xi_r-{1\over 6})R +{\lambda_r\over2}\phi_c^2+{Q\over C} \right) $$
$$\times ln \left[{m^2_r + (\xi_r-{1\over 6})R +{\lambda_r\over2}\phi_c^2+{Q\over C}\over m^2_r + Q/C} \right]]\eqno{(3.15)}$$
\\
Using this formula we can draw three conclusions.

    (1) In general, only for some suitable values of scalar-gravitational coupling $\xi_r$ mass $m_r$, curvature R and anisotropy Q could the symmetry be radiatively broken or restored.

   (2) As the second term, which is the quantum field correction term, is zero if $\xi_r=1/6$ and/or R =0, the results thus show that the fate of symmetry will not be changed in the cases of conformal coupling and/or vanishing scale curvature. This result has been reported by us in [ 12] in which the massless case was studied.

    (3) In the case of large anisotropy (under the conditions of $ Q>>\dot H +  H^2$ and $Q >> C m_r^2$), then equation (3.15) becomes

$$ m^2_{eff} \rightarrow  6\xi_r {Q\over C} + {\lambda_r\phi_c\over 32\pi^2}{6Q\over C}\left[\xi_r ln|6\xi_r| - \left(\xi_r-{1\over 6}\right)\right]\eqno{(3.16)}$$
\\
Although the second term, which is the quantum.field correction term, becomes negative if  $\xi_r << 0$, this does not mean that the symmetry will be in the broken phase for a $\phi^4$ theory in a spacetime with large anisotropy even the quantum-field effect has been introduced. The reason is that in the case of large anisotropy $W_k^2 \rightarrow  6\xi_r Q$ , thus the condition of a real function of $W_k$, will constrain $\xi_r$ to be a non-negative value. Note. that the measure of adiabatic $\dot W_k/ W_k^2$, which approaches ${1\over2}(6\xi_r)^{-1/2}\dot Q Q^{-1/2}$  in the case of large anisotropy, must be a small value in order to validate the WKB approximation we adopted. 

   Thus the problem of whether the symmetry will be restored through the quantum field effect for the $\phi^4$ theory in a spacetime with large anisotropy cannot be concluded from the method of WKB approximation. It falls short of the author's expectation. In fact, the primary motivation of this paper is to present an example which will, through calculating the effective potential, explicitly show that the larger spacetime anisotropy will restore the cosmological symmetry.

    Note that the temperature can be considered as a sort of spacetime anisotropy,
and cosmological symmetry will be restored in the high temperature. Note also that, although in [6] it had been shown that the radiative correction terms arising from the anisotropy of Bianchi type I spacetime are positive and proportional to the magnitude of anisotropy Q, we cannot thus claim that the cosmological symmetry will be restored for high Q value, because the calculation in [6] has used the small-Q-value assumption.

   We thus study the following case.

\subsection{ A Kasner spacetime}

We consider the Bianchi type I spacetime with a Kasner metric
$$ds^2 = dt^2 -t^2 dx^2 -dy^2-dz^2 = {2\over 3}\eta~ d\eta^2- ({2\over 3}\eta)^{3}dx^2 -dy^2-dz^2  \eqno{(3.17)}$$
where $\eta$ is the conformal time.  Fulling et at [17] have obtained the mode solution in the above spacetime

$$\chi_k = N\eta^{1/2} Z_{ik_1}\left[({2\over 3}\eta)^{3/2}(k^2+m^2+ {1\over2}\lambda_r\phi_c^2)^{1/2}\right], ~~~ k^2= k_2^2+k_3^2,  \eqno{(3.18)}$$
where $Z_{ik_1}$ is the Bessel function and the normalization constant N is determined to be $i\pi^{1/2}/2$ by the Wronskian condition (2.13). Fulling et at [17] have presented detailed analyses to show that the solution given in terms of the Hankel function , $H_{ik_1}^{(2)}$, will be positive frequency with respect to the adiabatic definition [16]. We will in this paper adopt their results to evaluate the integration in equation (3.4).

   It should be noted here that the metric in equation (3.17) can be reduced to the
Minkowski metric by a suitable transformation [ 17]; however, as soon as we enter into the realm of quantized field theory, these two systems of references will not be equivalent to each other. The reason is that for a quantum field in a curved spacetime it will have a time-dependence vacuum and thus the vacuum states based on the Minkowski metric and the original cosmological spacetime are not equivalent to each other, and the non-equivalent vacuum state will in general imply' different quantum field effects. In other words: although our model spacetime is in fact merely a coordinatization of Minkowski spacetime, it can lead to non-trivial quantum effects. Thus there exist papers which investigate some other quantum-field effects in the above spacetime [ 18, 19]. Our result
also presents a new quantum effect. 

    Note also that, in general, for a time-dependent background field the effective
potential is not well defined and one needs to work with the effective action instead.   The effective potential can be, at most, taken as the zeroth-derivative-order approximation to the effective action as a functional of the background field [8, 9].  We will leave the study of effective action to the further publication. Nevertheless, this does not say that the present calculation about the effective potential is meaningless. In fact, a chosen mode function in equation (3.18) will approach the WKB solution in the asymptotic region, and in this region the effective potential becomes a good approximation to the zeroth-derivative order the effective action in which we are interested.  In other words, our calculation on the effective potential can be used to investigate the quantum field effects on the cosmological phase transition, at least, in a suitable (asymptotical) region. The same argument can be applied to the mode studied in section 4.

   We now begin to perform the calculation, Using the integration representation of the HankeI function [20] we have the following relation

$$ \int dk_1 H_{ik_1}^{(2)}(x)^* H_{ik_1}^{(2)}(x) \hspace{11cm}$$
$$ = {4\over\pi}\int_{-\infty}^\infty dk_1 \int_1^\infty dt_1 \int_1^\infty dt_2 |\Gamma({1\over2}-ik_1)|^{-2} (t^2_1-1)^{-1/2-ik_1}(t^2_2-1)^{-1/2+ik_1} exp^{-ix(t_1-t_2)}$$
$$=8 \int_1^\infty dt_1 \int_1^\infty dt_2 {exp^{-ix(t_1-t_2)}\over[(t_1^2-1)(t_2^2-1)]^{1/2}} \int_0^\infty dk_1 sech(\pi k_1)cos\left(k_1 ln(t^2_1-1)^{-1}(t^2_2-1)\right) $$
$$= 8 \int_1^\infty dt_1 \int_1^\infty dt_2 ~~{exp^{-ix(t_1-t_2)}\over (t_1^2-1)+(t_2^2-1)} = 8 \int_2^\infty dy_2 \int_{-\infty}^\infty dy_1 ~~{exp^{-ixy_1)}\over y_1^2+y_2^2-4 } hspace{2cm}$$
$$= 8 \pi  \int_2^\infty dy_2 ~~(y^2_2-4)^{-1/2}exp[-x(y_2^2- 4)^{-1/2}] = 8 \pi  \int_0^\infty dw ~~(w^2+4)^{-1/2}exp(-xw) $$
$$= 4\pi^2[H_0(2x)-N_0(2x)] = 4\pi \sum_{n=0} {\Gamma (n+1/2) \over \Gamma (-n+1/2)}x^{-2n-1}, \hspace{4cm}\eqno{(3.19)}$$
\\
where $H_0$ is the Struve function and $N_0$ is the Neumann function [20]. Substituting the above result into equation (3.4) we have

$$ <\phi_q^2> = {3\over 8\pi^2}\int_0^\Lambda dk k \sum_{n=0} {\Gamma (n+1/2) \over \Gamma (-n+1/2)}\left[({2\over 3}\eta)^{3/2}(k^2+m^2+ {1\over2}\lambda_r\phi_c^2)\right]^{-n-1/2}$$
$$ =  {3\over 16\pi^2}\left[({2\over 3}\eta)^{3/2}[\Lambda - (m^2+ {1\over2}\lambda_r\phi_c^2)^{1/2}] + F(\eta) \right], \eqno{(3.20)}$$
\\
where 

$$ F(\eta) =  {\eta \over 8\pi^2} \sum_{n= 1} {\Gamma (n+1/2) \over \Gamma (-n+1/2)}({2\over 3}\eta)^{-3n-3/2} (m^2+ {1\over2}\lambda_r\phi_c^2)^{-n+1/2}.  \eqno{(3.21)}$$
\\
We have introduced a momentum cut-off $\Lambda$ to regularize the integration. Note that the series $F(\eta)$ is expanded in the parameter $[\eta^3 (m^2+ {1\over2}\lambda_r\phi_c^2)]^{-1}$ and thus the massless case is excluded.

    Substituting the above value into equation (2.7) and keeping only the leading term as $\eta \rightarrow \infty$ we then have a simple result

$${\partial V_{eff}\over \partial \phi_c} = (m^2_r +\delta m)\phi_c  + {\lambda_r\over3!}\phi_c^3 + {\lambda_r\phi_c\over 32\pi^2}~\eta ^{-3/2}\left[\Lambda - (m^2+ {\lambda_r\over2}\phi_c^2)^{1/2}\right]. \eqno{(3.22)}$$
\\
Notice that the curvature R is zero in the above Kasner spacetime and only the
counterterm $\delta m^2$  is needed in our model. The renormalization conditions equation (2.8) give
$$\delta m^2 = -{\lambda_r \over 32\pi^2}~\eta ^{-3/2} (\Lambda -m_r), \eqno{(3.23)}$$
and we finally obtain the derivative of the one-loop renormalized effective potential

$${\partial V_{eff} \over \partial\phi_c} = m^2_r\phi_c + {\lambda_r\over3!} \phi_c^3 -  {\lambda_r\phi_c\over 32\pi^2}~\eta ^{-3/2}\left[(m_r^2+ {\lambda_r\over2}\phi_c^2)^{1/2} - m_r \right].\eqno{(3.24)}$$

  We can see from the above equation that the second term, which is always negative, may be dominant for some large (but not very large) value of $\phi_c$. In this case, the effective mass becomes negative and thus the universe is in the symmetry-broken phase.  (In fact, in the Taylor expansion of the effective potential will appear a $\phi_c^3$ term, which indicates that there is a first-order phase transition [22].) However, as the time evolves the anisotropy of spacetime grows and the second term will approach zero by the factor $\eta^{-3/2}$, then the effective mass becomes positive (note that the above analyses require $m_r^2> 0$) and the universe will be restored to the symmetry phase. This means that even though the symmetry may be broken through the quantum field effect the
large anisotropy coming from the cosmological evolution will restore it to the symmetry phase.

\section {$\phi^4$ on Kantowski-Sachs spacetime}

We next consider the Kantowski-Sachs spacetime which has the line element
$$     ds^2=dt^2- X^2(t) dx^2- Y^2(t)\left[d\theta^2+sin^2 \theta  d\varphi^2\right].\eqno{(4.1)}$$
In this model spacetime the mode function and measure defined in equation (2.11) become

$$ Y_k(x) = (2\pi)^{-1/2} exp(i k x) Y_{lm}(\theta,\varphi), \eqno{(4.2)}$$
$$ \int d\mu(k) = \int dk \sum_{lm} , ~~~ l=0,1,...~~~ m= -l,....,l,  \eqno{(4.3)}$$
\\
where $ Y_{lm}(\theta,\varphi)$  is the spherical harmonics function.  Using the addition theorem of the spherical harmonics function equation (2.14) becomes

$$ <\phi^2_q(\eta)> = {1\over 8\pi^2 C(\eta)} \int d k \sum_l (2l+1) \chi_{k,l}(t) \chi^*_{k,l}(t),    ~~~~C(t) = (XY^2)^{1/2}.   \eqno{(4.4)}$$
\\
In the above equation the time-part mode function $\chi_{k,l}(t)$ satisfies the wave equation

$$\ddot \chi_{k,l} + \left[m^2_r + \xi_r R +{\lambda_r\over2}\phi_c^2+ {k^2\over X^2} + {l(l+1)\over Y^2} - {1\over4}(2\dot W + W^2) \right] \chi_{k,l} =0, \eqno{(4.5)}$$
\\
where the spacetime curvature function R and function W are
                                              
  $$  R=2\dot W + W^2+h_1^2+ 2h_2^2+ {2\over Y^2}, ~~~     h_1= \dot X/X, ~~~ h_2= \dot Y/Y,      \eqno{(4.6)}$$
$$                W=h_1+2 h_2.         \eqno{(4.7)}$$

   To evaluate $<\phi^2_q(\eta)>$ from equation (4.4) we need an explicit function form of $\chi_{k,l}$  thus let us consider the following two cases.

\subsection{WKB approximation}

When the metric is slowly varying equation (4.5) possesses the WKB approximation solution

$$\chi_{k,l} = (2W_k)^{-1/2} exp\left( -i\int d t ~W_k\right), \eqno{(4.8)}$$
\\
where $W_k$ can be approximated by [16,17]

$$ W_k = \left[m^2_r + \xi_r R +{\lambda_r\over2}\phi_c^2+ {k^2\over X^2} + {l(l+1)\over Y^2} - {1\over4}(2\dot W + W^2) \right]^{1/2} \eqno{(4.9)}$$
\\
Substituting the above solutions into equation (4.4) we have

$$<\phi_q^2> = {1\over 16\pi^2 C} \int dk \sum_{l} (2l+1)  \left[m^2_r + \xi_r R +{\lambda_r\over2}\phi_c^2+ {k^2\over X^2} + {l(l+1)\over Y^2} - {1\over4}(2\dot W + W^2) \right]^{-1/2}$$
$$=  {1\over 4\pi^2} \int_0^\Lambda \int_{1/2Y}^\Lambda  dz z  \left[m^2_r + \xi_r R +{\lambda_r\over2}\phi_c^2+ k^2 +z  - {1\over4}(2\dot W + W^2) \right]^{-1/2}$$
$$=  {1\over 8\pi^2}[[\sqrt2 -1+ln(\sqrt2 +1)]\Lambda ^2- [m^2_r + \xi_r R +{\lambda_r\over2}\phi_c^2 - {1\over4}(2\dot W + W^2)]$$
$$\times [{1\over2} +ln[2\Lambda/(\sqrt 2+1)] + {1\over2}[m^2_r + \xi_r R +{\lambda_r\over2}\phi_c^2 - {1\over4}(2\dot W + W^2)] $$
$$ \times ln[m^2_r + \xi_r R +{\lambda_r\over2}\phi_c^2 - {1\over4}(2\dot W + W^2)] - {1\over 4 Y^2} ln(\sqrt2 +2)],\eqno{(4.10)}$$
\\
where we have replaced the summation over t by an integration over $z (=(l+{1\over2})/Y)$ and introduced a momentum cut-off $\Lambda$ to regularize the $k$ and $l$ integration. From the renormalization conditions of equation (2.8) we can evaluate the renormalization counterterms

$$\delta m^2 = -{\lambda_r\over 16\pi^2} [[\sqrt2 -1+ln(\sqrt2 +1)]\Lambda^2 - [m^2_r - {1\over4}(2\dot W + W^2)]$$
$$\times [{1\over2} +ln[2\Lambda/(\sqrt 2+1)] + {1\over2}[m^2_r  - {1\over4}(2\dot W + W^2)] $$
$$ \times ln[m^2_r - {1\over4}(2\dot W + W^2)] - {1\over 4 Y^2} ln(\sqrt2 +2)],\eqno{(4.11)}$$

$$ \delta \xi = -{\lambda_r\over 16\pi^2}[-\xi_r  ln[2\Lambda/(\sqrt 2+1)] + {1\over2}\xi_r ln[m^2_r - {1\over4}(2\dot W + W^2)]] ,\eqno{(4.12)}$$

$$ \delta \lambda  = -{3\lambda_r^2 \over 16\pi^2} [-\lambda _r  ln[2\Lambda/(\sqrt 2+1)] + {1\over2}\lambda_r ln[m^2_r - {1\over4}(2\dot W + W^2)]]\eqno{(4.13)}$$
\\
Substituting these results into equation (2.7) we finally obtain the derivative of the one-loop renormalized effective potential

$$ {\partial V_{eff} \over \partial\phi_c} = m^2_r\phi_c + \left[1 - {\lambda_r\over32\pi^2}\right] \xi_r R \phi_c + \left({\lambda_r\over3!}-{\lambda_r^2\over 64\pi^2}\right)\phi_c^3 \hspace{5cm}$$
$$+{\lambda_r\phi_c\over 32\pi^2}  \left[m^2_r + \xi_r R +{\lambda_r\over2}\phi_c^2 - {1\over4} (2\dot W+W^2) \right] ln \left[{m^2_r + \xi_r R +{\lambda_r\over2}\phi_c^2 + {1\over4} (2\dot W+W^2) \over m^2_r - {1\over4} (2\dot W+W^2) }\right] \eqno{(4.14)}$$
\\

   We are now in a position to discuss the quantum field effects on the cosmological phase transition. Taking the derivation of the above equation, then in the limit  $\phi_c\rightarrow 0$ we obtain the effective mass

$$ m^2_{eff} = (m_r^2 +\xi_r R) + {\lambda_r\phi_c\over 32\pi^2}  \left[-\xi_r R + [m_r^2 +\xi_r - {1\over4} (2\dot W+W^2) \right] ln \left[{m^2_r + \xi_r R - {1\over4} (2\dot W+W^2) \over  m^2_r - {1\over4} (2\dot W+W^2) }\right] \eqno{(4.15)}$$
\\
Using this formula we can draw the following two conclusions.

   (1) In general, only for some suitable values of scalar-gravitational coupling $\xi_r$, mass $m_r^2$, curvature R and anisotropy Q could the symmetry be radiatively broken or restored.

   (2) As the second term, which is the quantum field correction term, is zero if $\xi_r=0$  (not the value of ${1/6}$ in the Bianchi type I model spacetime) and/or R =0, the results thus show that the fate of symmetry will not be changed in the cases of minimal coupling and/or vanishing scale curvature.

   Note that the above result is very like that in the Bianchi type I spacetime (see
equation (3.14)), although the topology of the Kantowski-Sachs spacetime is very
different from that of the Bianchi type I spacetime. 

   We also consider the next case in which the mode function has an exact form and we can obtain an exact value of $<\phi^2>$.

\subsection{A Kantowski-Sachs spacetime}

We consider a Kantowski-Sachs spacetime with a metric

$$ds^2 = dt^2- t^2 dx^2- Y_0^2\left[d\theta^2+sin^2 \theta  d\varphi^2\right]. \eqno{(4.16)}$$
\\
where $Y_0$ is a constant. The mode equation (4.5) in the above metric function has the solution

$$\chi_k = N t^{1/2} Z_{ik_1}\left[ t \left(m_r^2+ {\lambda_r\over2}\phi_c^2 +{l(l+1)\over Y^2_0}+{2\xi_r\over Y^2_0}\right)^{1/2}\right], \eqno{(4.17)}$$
\\
where $Z_{ik}$ is the Bessel function and the normalization constant N is determined to be $i\pi^{1/2}/2$ by the Wronskian condition (2.13). We can follow the analyses of Fulling (2) et al [ 17] and show that the solution given in terms of the Hankel function $H_{ik}^{(2)}$ will be positive frequency with respect to the adiabatic definition [21]. Substituting the mode solution in the above equation into equation (4.4) and performing the k integration just like that in equation (3.19), we then obtain

$$ <\phi_q^2> = {1\over 8\pi} \int_0^\Lambda dx \left[H_0\left[ 2t\left(m_r^2+ {1\over2}\lambda_r\phi_c^2 + x+{2\xi_r\over Y^2_0}\right)^{1\over2}\right] - N_0\left[ 2t\left(m_r^2+ {1\over2}\lambda_r\phi_c^2 + x+{2\xi_r\over Y^2_0}\right)^{1\over2}\right]\right]$$
$$= {1\over 4\pi^2}\int_0^\Lambda dx \sum_{n=0} {\Gamma (n+1/2) \over \Gamma (-n+1/2)}\left[2 t \left(m_r^2+ {1\over2}\lambda_r\phi_c^2+x+{2\xi_r\over Y^2_0}\right)^{1/2}\right]^{-n-1/2}$$

$$ =  {1\over 4\pi^2}\left[\Lambda^{1/2} - \left(m^2+ {1\over2}\lambda_r\phi_c^2 + {2\xi_r\over Y^2_0}\right)^{1/2}  \right] +O(t^{-3}), \hspace{3cm}\eqno{(4.18)}$$
\\
where we have replaced the summation over t by an integration over $x(=(l(l+ 1)/Y_0^2))$  and introduced a momentum cut-off $\Lambda$  in the x integration. From the renormalization conditions of equation (2.8) we can evaluate the renormalization counterterms

$$\delta m^2 =  -{\lambda_r\over 8\pi^2 t }\left[\Lambda^{1/2} - \left(m^2+  {2\xi_r\over Y^2_0}\right)^{1/2}  \right], \eqno{(4.19)}$$
\\
and we finally obtain the derivative of the one-loop renormalized effective potential

$${\partial V_{eff} \over \partial\phi_c} = m^2_r\phi_c + \xi_r R \phi_c +{\lambda_r\over3!}\phi_c^3 + {\lambda_r \phi_c\over 8\pi^2 t }\left[ \left(m^2+  {2\xi_r\over Y^2_0}\right)^{1/2}-\left(m^2+ {1\over2}\lambda_r\phi_c^2 + {2\xi_r\over Y^2_0}\right)^{1/2}\right] + O(t^{-3}). \eqno{(4.20)}$$
\\
    The result is very like that in the Kasner spacetime (note that the topology of these two spacetimes is different, for example, the curvature of Kasner space we consider is vanishing while that of the Kantowski-Bachs space is finite) and we have the following same conclusion: the universe may be in the symmetry-broken phase through a first-order phase transition, however, as the time evolves the anisotropy of spacetime grows and the universe will always be restored to the symmetry phase.

\section{ Discussions}

After the new inflationary-universe scenario [23] is proposed, it is found by many
authors that the gravitational effect plays an important role in the cosmological phase transition [2-I4]. From their analyses about the loop correction effective potential one then concludes that the scalar-gravitational coupling $\xi$ and the magnitude of the scalar curvature R crucially determine the fate of symmetry.

    However, due to the mathematical technical difficulty, literature concerning the
analyses on the anisotropic spacetime is very limited [4-6, 12]. In this paper we have presented such analyses in some types of anisotropic spacetime.

    We have evaluated the one-loop renormalized effective potentials for the massive $\phi^4$ theory on the spatially homogeneous models of Bianchi type I and Kantowski-Sachs type. To enable the analytic results to be found we assume that the spacetime metric is either slowly varying or has a specified form. 

    In the case of the slowly varying metric we find the following conclusions: (1) In general, only for some suitable values of scalar-gravitational coupling $\xi$, mass $m$, curvature R and anisotropy Q could the symmetry be radiatively  broken or restored.  (2) The fate of symmetry will not be changed in the cases of conformal coupling for Bianchi type I space and minimal coupling for the Kantowski-Sachs, and/or vanishing scale curvature for both cases.

    We have also to consider a specified Bianchi type I space and a specified Kantowski-Sachs space. We have argued that although the specified Bianchi type I metric can be reduced to the Minkowski metric by a suitable transformation, however, as soon as we enter into the realm of quantized field theory these two systems of reference will not be equivalent to each other because the vacuums in the two systems are different.  Thus our result can offer a nontrivial quantum effect. 

   We also note that for a time dependent background field the effective potential is not well defined and one needs to work with the effective action instead. However, our calculation about the effective potential can be used to investigate the quantum field effects on the cosmological phase transition, at least, in a suitably asymptotic region.

    For both the specified spacetimes, our results indicate that they have the same
phase transition although they have very different spacetime topology. The result shows that the universe may be in the symmetry-broken phase through a first-order phase transition, however, as the time evolves the anisotropy of spacetime grows and the universe will be restored to the symmetry phase.

    Finally, we hope that our results may be helpful in our understanding of quantum field effects on the cosmological phase transition in anisotropic spacetimes.

\newpage
\begin{enumerate}
\item  Birrell N D and Davies PC W 1982 Quantum Fields in Curved Space (Cambridge: Cambridge University Press)
\item  Hu B L 1983 Proc. tOth Int. Conf. on General Relativity and Gravitation ed B Bertotfi, F deFelice and A Pascolini (Rome: Consiglio Nazionale Delle ~cerche) p 1086
\item Abbott L F 1981 Nucl. Phys. B 185 233\\
      Ford H and Toms D J I982 Phys. Rev. D 25 1510\\
      Allen B 1983 Nuct. Phys. B 226 228; 1983 Ann. Phys., NY 161 152\\
      Vilenkin A 1983 Nuct. Phys. B 226 504
\item Critchley R and Dowker J S 1982 J. Phys. A: Math. Gen. 15; 157
\item Shen T C, Hu B L and O'Connor D J 1985 Phys. Rev. D 3! 2401
\item Futamase T 1984 Phys. Rev. D 29 2789
\item Hu B L and O'Connor D J 1984 Phys. Ret.,. D 30 74; 1987 D 34 2535; 1987 D 36 1701\\
        Hu B L and Zhang Y 1988 Phys. Rev. D 37 2125
\item  Hu B L 1983 Phys. Lett. 123B 189\\
         Chen L F and Hu B L 1985 Phys. Lett. 160B 36
\item  Sinha S and Hu B L 1988 Phys. Rev. D 38 2423
\item  Ringwald A 1987 Phys. Ret,. D 36 2598; 1987 Ann. Phys., NY 177 129
\item Chimento L P, Jakubi A S and Pullin J 1989 Class. Quantum Grav. 6.45
\item  Huang W H 1990 "Quantum Field Effect on Symmetry Breaking and Restoration in Anisotropic Spacetimes", Phys. Rev. D 42 1287 [gr-qc/0401037]
\item  Huang W H 1991 ""Coleman Weinberg Symmetry Breaking in the Early Universe with an Inhomogeneity "",  Class. Quantum Grav. 8 83 
\item  Huang W H 1991 "Finite-Temperature Cosmological Phase Transition in a
Rotating Spacetime", Class. Quantum Grav. 8 1471 [gr-qc/0401042]
\item  Zel'dovich Ya B and Starobinsky A A 1972 Zh. Eksp, Teor. Fiz, 61 2162 (Soy. Phys.-JETP 34 1159)
\item  Parker L and Fulling S A 1974 Phys, Rev, D 9 341
\item  Fulling S A, Parker L and Hu B L 1974 Phys. Rev. D 10 3905
\item  Nariai H 1976 Nuovo Cimento 35 259; 1977 :Prog. Theor. Phys. 57 67
\item  Berger B K 1975 Phys, Rev, D 12 368
\item  Gradshteyn I S and Ryzhik I M 1980 Table of Integrals, Series, and Products (New York: Academic)
\item  Huang W H 1989 " Particle Creation in Kaluza-Klein Cosmology" 
Phys. Lett, 140A 280 [gr-qc/0308086]
\item  Stanley H E 1971 Introduction to Phase Transition and Critical Phenomena (Oxford: Oxford University Press)
\item  Guth A 1981 Phys. Rev. D 23 347\\
      Langacker P 1981 Phys. Rep. 72 185\\
      Linde A D 1982 Phys. Lett. 108B 489\\
      Albrecht A and Steinhardt P J 1982 Phys. Rev. Left. 48 1220
\end{enumerate}
\end{document}